\documentclass[12pt]{article}

           \usepackage{amsfonts}
           \usepackage{amssymb}

\def\be{\begin{eqnarray}}
\def\ee{\end{eqnarray}}
\def\bee{\begin{eqnarray*}}
\def\eee{\end{eqnarray*}}

\newtheorem{thm}{Theorem}

\newtheorem{conj}[thm]{Conjecture}

\newtheorem{defn}[thm]{Definition}

           \def\tr{\hbox{Tr}}

\def\bra{\langle}
\def\ket{\rangle}

\def\ot{\otimes}

             \title{
An application of a matrix inequality in quantum information theory}
\author{Christopher King
\\ Department of Mathematics
\\ Northeastern University
\\ Boston MA 02115
\\
{\normalsize king@neu.edu}
}
\begin{document}

\maketitle

\begin{abstract}
Quantum information theory has generated several interesting conjectures
involving products of completely positive maps on matrix algebras, also
known as quantum channels.
In particular it is conjectured that the output state
with maximal $p$-norm from a product channel is always a product state.
It is shown here that the Lieb-Thirring inequality 
can be used to prove
this conjecture for one special case, namely when 
one of the components of the product channel is
of the type known as a diagonal channel.
\end{abstract}

\pagebreak

%\tableofcontents

%\bigskip

\section{Introduction}
The minimal output entropy of a quantum channel $\Phi$ is defined by
\be
S_{\rm min}(\Phi) = \inf_{\rho} S(\Phi(\rho))
\ee
where $S$ is the von Neumann entropy, and the $\inf$ runs over states
in the domain of $\Phi$. The following
additivity  property is conjectured.
\begin{conj}\label{conj1}
Let $\Phi$ and $\Psi$ be any quantum channels, that is completely 
positive, trace-preserving maps on
finite-dimensional matrix algebras. Then
\be\label{min.add}
S_{\rm min}(\Phi \ot \Psi) =  S_{\rm min}(\Phi) +  S_{\rm min}(\Psi)
\ee
\end{conj}
\medskip

Within the last year, Shor \cite{Shor2} proved that Conjecture 
\ref{conj1} is equivalent to several other
outstanding conjectures in quantum information theory, among them 
additivity of Holevo capacity
of a quantum channel, and additivity of the entanglement of 
formation. Thus a proof of Conjecture \ref{conj1} would settle quite 
a few outstanding problems in the field.

\medskip
The additivity conjecture has been proved for several special
classes of channels \cite{Shor1}, \cite{Ki3}, \cite{Ki4}.
The Lieb-Thirring inequality
\cite{LT} was a key ingredient in several of those proofs.
The purpose of this paper is to show how 
the Lieb-Thirring inequality can be used to 
demonstrate (\ref{min.add}) for another class known as
the `diagonal' channels.
\medskip

The 
problem is attacked by making use  of the maximal output $p$-norm for 
$p \geq 1$, also called the maximal output purity  of a channel \cite{AHW}, which 
is defined by
\be\label{def:nu}
{\nu}_p(\Phi) = \sup_{\rho} || \Phi(\rho) ||_p = \sup_{\rho} \bigg(
\tr \Big( \Phi(\rho) \Big)^p \bigg)^{1/p}
\ee
The derivative of ${\nu}_p(\Phi)$ at $p=1$ is the negative minimal 
output entropy, so Conjecture
\ref{conj1} is a consequence of the following stronger conjecture.

\begin{conj}\label{conj2}
There is some $p_0 > 1$, such that for all
  quantum channels $\Phi$ and $\Psi$, and all $1 \leq p \leq p_0$,
\be\label{p.mult}
{\nu}_p(\Phi \ot \Psi) =  {\nu}_p(\Phi) \,\, {\nu}_p(\Psi)
\ee
\end{conj}
\medskip
In this paper we consider a class of channels known as the
`diagonal' channels, and show how the Lieb-Thirring inequality
can be used to derive (\ref{p.mult}) under the assumption that at 
least one of the channels
$\Phi$ or $\Psi$ is in this class.
It turns out that the multiplicativity result for diagonal channels holds for all
$p \geq 1$, so this might lead one to hope that $p_0 = \infty$ in 
Conjecture \ref{conj2}.
However it is known that multiplicativity fails in general for $p 
\geq 5$ \cite{WH},
and indeed this probably provides evidence that the strategy used in 
this paper cannot
be directly extended to prove Conjecture \ref{conj2} in the general case.
Nevertheless it seems worthwhile to explain the approach used, as
the techniques may be useful for other reasons, and the results may
have other applications in quantum information theory. The method of proof is
quite similar to the approach used by the author to prove Conjecture 
\ref{conj2} for the class of entanglement-breaking channels
\cite{Ki2}.

\section{Statement of results}
Since we will be concerned with $p$-norms from now on, the 
trace-preserving condition for quantum channels is unimportant, and 
so we deal instead with completely positive maps.
The diagonal class of channels was described by Landau and Streater \cite{LS}.
Recall that the Hadamard product of two $n \times n$ matrices $A$ and 
$B$  is defined by
\be
(A * B)_{ij} = A_{ij} B_{ij}
\ee

\begin{defn}
The CP map $\Phi$ is called diagonal if there is a positive semidefinite matrix
$C$ such that
\be\label{def:diag}
\Phi(\rho) =  C * \rho
\ee
\end{defn}
\medskip
If $C = | \psi \ket  \bra \psi | $ is rank one, then (\ref{def:diag}) 
can be written
\be
C * \rho = {\rm Diag}( \psi) \, \rho \, {\rm Diag}( \psi)^{*}
\ee
where ${\rm Diag}( \psi)$ is the diagonal $n \times n$ matrix with 
the components of
$| \psi \ket $ along the diagonal. Using the spectral representation 
it follows that
a map is diagonal if and only if it has a Kraus representation with 
all diagonal
matrices.

\medskip
Our main result is stated below in Theorem \ref{thm1}.

\begin{thm}\label{thm1}
Let $\Phi$ be  a diagonal map, and let $\Psi$ be any
other CP map. Then for all $p \geq 1$,
\be
{\nu}_p (\Phi \ot \Psi) = {\nu}_p (\Phi) \,\, {\nu}_p (\Psi)
\ee
\end{thm}

\medskip
The main tool used in the proof is the Lieb-Thirring inequality \cite{LT},
which we now state. 
Let $K \geq 0$ be a
positive semidefinite $n \times n$ matrix, and let $V$ be any $k \times n$ 
matrix. Then for all
$p \geq 1$,
\be\label{L-T}
\tr \Big( V K V^{*} \Big)^p \leq \tr (V^{*} V)^{p/2} \, K^p \, (V^{*} 
V)^{p/2} 
= \tr (V^{*} V)^p \, K^p
\ee
There are several proofs of this inequality \cite{LT}, \cite{Ar}. The 
original proof of Lieb and Thirring
employs Epstein's concavity theorem \cite{Ep}, which is based on a 
combination of spectral
theory and analytic continuation methods.
\medskip

\section{The factorization}
The goal of this section is to rewrite the output of the product channel
$\Phi \ot \Psi$ in the factorized form $V K V^{*} $ so that (\ref{L-T}) can be applied.
We assume that $\Phi$ is a diagonal channel which acts by
Hadamard product with the $n \times n$ matrix $C$.
Let $\rho$ be a state on ${\bf C}^{kn}$, that is a positive semidefinite 
${kn \times kn}$ matrix with trace 1, for some $k \geq 1$. Then
$\rho$ can be written as a $n \times n$ block matrix where the blocks 
$(\rho)_{ \,ij}$ are
$k \times k$ matrices. The diagonal blocks  $(\rho)_{ \,ii}$ are 
positive semidefinite,
and we define
\be
\alpha_i = \tr (\rho)_{ \,ii}
\ee
Define a new $kn \times kn$ matrix $\tau$ with blocks
\be
(\tau)_{ \,ij} = (\alpha_i \alpha_j)^{-1/2} \,\, (\rho)_{ \,ij}
\ee
and let $A$ denote the $n \times n$ matrix with entries $A_{ij} = 
(\alpha_i \alpha_j)^{1/2}$.
Then $\rho$ can be written as a Hadamard product of $A$ with
$\tau$, that is
\be
\rho = (A \ot J_k )*\tau
\ee
where $J_k$ is the $k \times k$ matrix with all entries equal to 1:
\be\label{def:J}
J_k = \pmatrix{1 & \dots & 1 \cr
\vdots && \vdots \cr
1 & \dots & 1}
\ee
Notice that $J_k$ acts as the identity for the Hadamard product.
Furthermore this decomposition commutes with the action of $\Psi$ on 
the second factor,
that is
\be\label{i-Psi}
(I \ot \Psi)(\rho) = (A \ot J_k)* \Big((I \ot \Psi)(\tau)\Big)
\ee
The map $\Phi \ot I$ acts on (\ref{i-Psi}) by a Hadamard product with 
the matrix $C$ on the first
factor.
This Hadamard product acts just on the matrix $A$, and the result is
\be\label{fact.A}
(\Phi \ot \Psi)(\rho) = \Big(\Phi(A) \ot J_k \Big)* \Big((I \ot \Psi)(\tau)\Big)
\ee
The next step is to factorize the matrix $(I \ot \Psi)(\tau)$. To do this,
let $V_1, \dots, V_n$ be the $k \times kn$ matrices which are the
block-rows of its square root, that is
\be
\Big({(I \ot \Psi)(\tau)}\Big)^{1/2} = \pmatrix{V_1 \cr \vdots \cr V_n}
\ee
Then it follows that
\be\label{fact1.Psi}
(I \ot \Psi)(\tau)  =  \pmatrix{V_1 \cr \vdots \cr V_n} 
\pmatrix{V_{1}^{*} & \dots & V_{n}^{*}} =
\pmatrix{V_1 V_{1}^{*} &  \dots & V_1 V_{n}^{*} \cr
\vdots & \ddots & \vdots \cr
V_n V_{1}^{*} & \dots  & V_n V_{n}^{*} \cr}
\ee
Notice that the diagonal terms are $\Psi((\tau)_{ \,ii}) = V_{i} 
V_{i}^{*}$, and since
$(\tau)_{ \,ii}$ is positive semidefinite with $\tr (\tau)_{ \,ii} = 1$ it 
follows that
\be\label{p-norm-V}
|| V_{i} V_{i}^{*} ||_p \leq {\nu}_{p}(\Psi)
\ee

Applying  the factorization (\ref{fact1.Psi}) to (\ref{fact.A}) gives
\be\label{fact1.A}
(\Phi \ot \Psi)(\rho) = \pmatrix{\Phi(A)_{\,11} V_1 V_{1}^{*} &  
\dots & \Phi(A)_{\, 1n} V_1 V_{n}^{*}
\cr
\vdots & \ddots & \vdots \cr
\Phi(A)_{\, n1} V_n V_{1}^{*} & \dots  & \Phi(A)_{\,nn} V_n V_{n}^{*} \cr}
\ee
Now the right side of (\ref{fact1.A}) can be rewritten as a product of three matrices:
\be\label{fact2.Psi}
\pmatrix{V_1 & 0 & \dots & 0 \cr
0 & V_2 & \dots & 0 \cr
\vdots && \ddots & \vdots \cr
0 & 0 & \dots & V_n \cr} \,
\pmatrix{\Phi(A)_{\,11} I' & \dots & \Phi(A)_{\,1n} I' \cr
\vdots & & \vdots \cr
\Phi(A)_{\,n1} I' & \dots & \Phi(A)_{\,nn} I'} \,
\pmatrix{V_{1}^{*} & 0 & \dots & 0 \cr
0 & V_{2}^{*} & \dots & 0 \cr
\vdots && \ddots & \vdots \cr
0 & 0 & \dots  & V_{n}^{*}}
\ee
where the $I'$ in the middle term is the $k n \times k n$ identity matrix.
This is the same as
\be\label{fact3.Psi}
\pmatrix{V_1 & 0 & \dots & 0 \cr
0 & V_2 & \dots & 0 \cr
\vdots && \ddots & \vdots \cr
0 & 0 & \dots  & V_n \cr} \,
(\Phi(A) \ot I') \,
\pmatrix{V_{1}^{*} & 0 & \dots & 0 \cr
0 & V_{2}^{*} & \dots & 0 \cr
\vdots && \ddots & \vdots \cr
0 & 0 & \dots  & V_{n}^{*}}
\ee
Therefore (\ref{fact1.A}) has been written in the factorized form
\be\label{factor}
(\Phi \ot \Psi)(\rho) = V \, K \, V^{*}
\ee
where $V$ is the $kn \times kn^2$ matrix
\be\label{def:bigV}
V = \pmatrix{V_1 & 0 & \dots & 0 \cr
0 & V_2 & \dots & 0 \cr
\vdots && \ddots & \vdots \cr
0 & 0 & \dots  & V_n}
\ee
and
\be
K = (\Phi(A) \ot I')
\ee

\section{Applying the inequality}
The last step is to apply the Lieb-Thirring inequality (\ref{L-T}) to (\ref{factor}).
It follows from (\ref{def:bigV}) that $\Big( V^{*} V \Big)^p$ is block diagonal, that is
\be
\Big( V^{*} V \Big)^p =
\pmatrix{(V_{1}^{*} V_1)^p & 0 & \dots & 0 \cr
0 & (V_{2}^{*} V_2)^p & \dots & 0 \cr
\vdots && \ddots & \vdots \cr
0 & 0 & \dots  & (V_{n}^{*} V_n)^p}
\ee
Also $K^p = (\Phi(A))^p \ot I'$, so the diagonal blocks of $K^p$ are just the
diagonal entries of $(\Phi(A))^p$ multiplied by the identity matrix $I'$. Hence
\be\label{L-T.d1}
\tr \Big( V^{*} V \Big)^p K^p =
\sum_{i=1}^n \tr (V_{i}^{*} V_i)^p \,\, \Big( (\Phi (A) )^p \Big)_{ \,ii}
\ee
The matrices $V_{i}^{*} V_i$ and $V_i V_{i}^{*}$ share the same nonzero spectrum,
and so (\ref{p-norm-V}) can be used to bound the terms $\tr (V_{i}^{*} V_i)^p$
on the right side of (\ref{L-T.d1}). This gives
\be
\tr \Big( V^{*} V \Big)^p K^p & \leq &
\sum_{i=1}^n \Big( {\nu}_{p}(\Psi) \Big)^p \Big( (\Phi (A) )^p \Big)_{ii} \\
& = &  \Big( {\nu}_{p}(\Psi) \Big)^p \, \tr (\Phi (A) )^p
\ee
Furthermore since $\rho$ is a state, it follows that
$\tr A = \tr \rho = 1$, and hence $\tr (\Phi (A) )^p$ can be bounded using the
definition (\ref{def:nu}). Putting it all together we deduce
\be
\tr (\Phi \ot \Psi)(\rho)^p \leq \Big( {\nu}_{p}(\Psi) \Big)^p \,
\Big( {\nu}_{p}(\Phi) \Big)^p
\ee
From this it follows that 
\be
{\nu}_p(\Phi \ot \Psi) \leq {\nu}_p(\Phi) \, {\nu}_p(\Psi)
\ee
The inequality in the other direction follows easily by restricting to product states,
hence Theorem \ref{thm1} is proved.

\bigskip
{\bf Acknowledgements}
This work was supported in part by
National Science Foundation Grant DMS--0101205. The author is
grateful to E. Lieb and M. B. Ruskai for first demonstrating that the
Lieb-Thirring inequality could be used to address the additivity problem,
and for allowing their work to be included in the Appendix of the paper
\cite{Ki1}.

{~~}

\end{document}